\documentstyle[12pt]{article}

\skewchar\fivmi='177
\skewchar\sixmi='177
\skewchar\sevmi='177
\skewchar\egtmi='177
\skewchar\ninmi='177
\skewchar\tenmi='177
\skewchar\elvmi='177
\skewchar\twlmi='177
\skewchar\frtnmi='177
\skewchar\svtnmi='177
\skewchar\twtymi='177
\def\@magscale#1{ scaled \magstep #1}
\skewchar\fivsy='60
\skewchar\sixsy='60
\skewchar\sevsy='60
\skewchar\egtsy='60
\skewchar\ninsy='60
\skewchar\tensy='60
\skewchar\elvsy='60
\skewchar\twlsy='60
\skewchar\frtnsy='60
\skewchar\svtnsy='60
\skewchar\twtysy='60

\catcode`@=11
\def\un#1{\relax\ifmmode\@@underline#1\else
        $\@@underline{\hbox{#1}}$\relax\fi}
\catcode`@=12

\def\a{\alpha}
\def\b{\beta}

\def\d{\delta}
\def\e{\epsilon}

\def\g{\gamma}

\def\l{\lambda}

\def\q{\theta}
\def\r{\rho}
\def\s{\sigma}

\def\z{\zeta}
\def\D{\Delta}

\def\O{\Omega}

\def\S{\Sigma}

\def\dslash{\not{\hbox{\kern-2pt $\partial$}}}
\def\Dslash{\not{\hbox{\kern-4pt $D$}}}
\def\pslash{\not{\hbox{\kern-2.3pt $p$}}}
 \newtoks\slashfraction
 \slashfraction={.13}
 \def\slash#1{\setbox0\hbox{$ #1 $}
 \setbox0\hbox to \the\slashfraction\wd0{\hss \box0}/\box0 }

\font\ro=cmsy10                          % font with rope
\def\kcr{{\hbox{\ro \char'170}}}                % right-handed rope
\def\ktl{{\hbox{\ro \char'170}}}        % top end for left-handed rope
\def\ktr{{\hbox{\ro \char'170}}}        % " right
\def\kbl{{\hbox{\ro \char'170}}}        % " bottom left
\def\kbr{{\hbox{\ro \char'170}}}        % " right
\def\plpl{\raise-2pt\hbox{$\raise3pt\hbox{$_+$}\hskip-6.67pt\raise0.0pt
\hbox{$^+$}\hskip 0.01pt$}}
\def\mimi{\raise-2pt\hbox{$\raise3pt\hbox{$_-$}\hskip-6.67pt\raise0.0pt
\hbox{$^-$}\hskip 0.01pt$}}

\def\bo{{\raise.15ex\hbox{\large$\Box$}}}               % D'Alembertian
\def\pa{\partial}                                       % curly d
\def\TH{{\raise.2ex\hbox{$\displaystyle \bigodot$}\mskip-4.7mu \llap H \;}}
\def\face{{\raise.2ex\hbox{$\displaystyle \bigodot$}\mskip-2.2mu \llap {$\ddot
        \smile$}}}                                      % happy face
\def\dvm{\raisebox{-.45ex}{\rlap{$=$}} }
\def\DM{{\scriptsize{\dvm}}~~}

\def\lin{\vrule width0.5pt height5pt depth1pt}
\def\dpx{{{ =\hskip-3.75pt{\lin}}\hskip3.75pt }}
\def\pp{{\mathchoice
          {%displaystyle
              \kern 1pt%
              \raise 1pt
              \vbox{\hrule width5pt height0.4pt depth0pt
                    \kern -2pt
                    \hbox{\kern 2.3pt
                          \vrule width0.4pt height6pt depth0pt
                          }
                    \kern -2pt
                    \hrule width5pt height0.4pt depth0pt}%
                    \kern 1pt
           }
            {%textstyle
              \kern 1pt%
              \raise 1pt
              \vbox{\hrule width4.3pt height0.4pt depth0pt
                    \kern -1.8pt
                    \hbox{\kern 1.95pt
                          \vrule width0.4pt height5.4pt depth0pt
                          }
                    \kern -1.8pt
                    \hrule width4.3pt height0.4pt depth0pt}%
                    \kern 1pt
            }
            {%scriptstyle
              \kern 0.5pt%
              \raise 1pt
              \vbox{\hrule width4.0pt height0.3pt depth0pt
                    \kern -1.9pt  %[e]=0.15pt
                    \hbox{\kern 1.85pt
                          \vrule width0.3pt height5.7pt depth0pt
                          }
                    \kern -1.9pt
                    \hrule width4.0pt height0.3pt depth0pt}%
                    \kern 0.5pt
            }
            {%scriptscriptstyle
              \kern 0.5pt%
              \raise 1pt
              \vbox{\hrule width3.6pt height0.3pt depth0pt
                    \kern -1.5pt
                    \hbox{\kern 1.65pt
                          \vrule width0.3pt height4.5pt depth0pt
                          }
                    \kern -1.5pt
                    \hrule width3.6pt height0.3pt depth0pt}%
                    \kern 0.5pt%}
            }
        }}

\def\sp#1{{}^{#1}}                              % superscript (unaligned)
\def\Tilde#1{\widetilde{#1}}                    % big tilde
\def\Hat#1{\widehat{#1}}                        % big hat
\def\Bar#1{\overline{#1}}                       % big bar
\def\leftrightarrowfill{$\mathsurround=0pt \mathord\leftarrow \mkern-6mu
        \cleaders\hbox{$\mkern-2mu \mathord- \mkern-2mu$}\hfill
        \mkern-6mu \mathord\rightarrow$}
\def\dvec#1{\vbox{\ialign{##\crcr
        \leftrightarrowfill\crcr\noalign{\kern-1pt\nointerlineskip}
        $\hfil\displaystyle{#1}\hfil$\crcr}}}           % <--> accent
\def\frac#1#2{{\textstyle{#1\over\vphantom2\smash{\raise.20ex
        \hbox{$\scriptstyle{#2}$}}}}}                   % fraction
\def\sfrac#1#2{{\vphantom1\smash{\lower.5ex\hbox{\small$#1$}}\over
        \vphantom1\smash{\raise.4ex\hbox{\small$#2$}}}} % alternate fraction
\def\bfrac#1#2{{\vphantom1\smash{\lower.5ex\hbox{$#1$}}\over
        \vphantom1\smash{\raise.3ex\hbox{$#2$}}}}       % "
\def\afrac#1#2{{\vphantom1\smash{\lower.5ex\hbox{$#1$}}\over#2}}    % "
\newskip\humongous \humongous=0pt plus 1000pt minus 1000pt
\def\caja{\mathsurround=0pt}
\def\eqalign#1{\,\vcenter{\openup2\jot \caja
        \ialign{\strut \hfil$\displaystyle{##}$&$
        \displaystyle{{}##}$\hfil\crcr#1\crcr}}\,}
\newif\ifdtup

\def\ref#1{$\sp{#1)}$}

\parskip=\medskipamount                 % space between paragraphs (LaTeX)
\thicklines                         % thick straight lines for pictures (LaTeX)

\def\oldheadpic{                                % old UM heading
        \setlength{\unitlength}{.4mm}
        \thinlines
        \par
        \begin{picture}(349,16)
        \put(325,16){\line(1,0){4}}
        \put(330,16){\line(1,0){4}}
        \put(340,16){\line(1,0){4}}
        \put(335,0){\line(1,0){4}}
        \put(340,0){\line(1,0){4}}
        \put(345,0){\line(1,0){4}}
        \put(329,0){\line(0,1){16}}
        \put(330,0){\line(0,1){16}}
        \put(339,0){\line(0,1){16}}
        \put(340,0){\line(0,1){16}}
        \put(344,0){\line(0,1){16}}
        \put(345,0){\line(0,1){16}}
        \put(329,16){\oval(8,32)[bl]}
        \put(330,16){\oval(8,32)[br]}
        \put(339,0){\oval(8,32)[tl]}
        \put(345,0){\oval(8,32)[tr]}
        \end{picture}
        \par
        \thicklines
        \vskip.2in}
\def\oldtitle#1#2#3#4{\oldheadpic\begin{center}\vglue.5in{\large\bf #1}\\[.6in]
        {#2}\\[.1in] {\it Department of Physics and Astronomy}\\
        {\it University of Maryland, College Park, MD 20742}\\[.6in]
        Physics Publication \#{#3}\\ {#4}\\[1.5in] {\bf ABSTRACT}\\[.1in]
        \end{center} \begin{quotation}}                 % old title stuff
\def\oldTitle#1#2#3#4#5#6#7{\oldheadpic\begin{center} \vglue .4in
        {\large\bf #1}\\[.4in]
        {#2}\\[.1in] {\it Department of Physics and Astronomy}\\
        {\it University of Maryland, College Park, MD 20742}\\[.1in]
        {#3}\\[.1in] {\it {#4}}\\ {\it {#5}}\\[.4in]
        Physics Publication \#{#6}\\ {#7}\\[.5in] {\bf ABSTRACT}\\[.1in]
        \end{center} \begin{quotation}}                 % " for 2 authors
\def\border{                                            % border
        \setlength{\unitlength}{1mm}
        \newcount\xco
        \newcount\yco
        \xco=-21
        \yco=12
        \begin{picture}(140,0)
        \put(\xco,\yco){$\ktl$}
        \advance\yco by-1
        {\loop
        \put(\xco,\yco){$\kcr$}
        \advance\yco by-2
        \ifnum\yco>-240
        \repeat
        \put(\xco,\yco){$\kbl$}}
        \xco=158
        \yco=12
        \put(\xco,\yco){$\ktr$}
        \advance\yco by-1
        {\loop
        \put(\xco,\yco){$\kcr$}
        \advance\yco by-2
        \ifnum\yco>-240
        \repeat
        \put(\xco,\yco){$\kbr$}}
        \put(-20,13){\tiny University of Maryland Elementary Particle
Physics University of Maryland Elementary Particle Physics University of
Maryland Elementary Particle Physics}
        \put(-20,-241.5){\tiny University of Maryland Elementary
Particle Physics University of Maryland Elementary Particle Physics
University of Maryland Elementary Particle Physics}
        \end{picture}
        \par\vskip-8mm}
\def\bordero{                                           % alternate border
        \setlength{\unitlength}{1mm}
        \newcount\xco
        \newcount\yco
        \xco=-31
        \yco=12
        \begin{picture}(140,0)
        \put(\xco,\yco){$\ktl$}
        \advance\yco by-1
        {\loop
        \put(\xco,\yco){$\kclr}
        \advance\yco by-2
        \ifnum\yco>-240
        \repeat
        \put(\xco,\yco){$\kbl$}}
        \xco=151
        \yco=12
        \put(\xco,\yco){$\ktr$}
        \advance\yco by-1
        {\loop
        \put(\xco,\yco){$\kcr$}
        \advance\yco by-2
        \ifnum\yco>-240
        \repeat
        \put(\xco,\yco){$\kbr$}}
        \put(-20,12){\ooo
bacdefghidfghghdhededbihdgdfdfhhdheidhdhebaaahjhhdahba

hgdedge
   hgfdiehhgdigicba}
        \put(-20,-241.5){\ooo
ababaighefdbfghgeahgdfgafagihdidihiidhiagfedhadbfd

ecdcdfa
   gdcbhaddhbgfchbgfdacfediacbabab}
        \end{picture}
        \par\vskip-8mm}
\def\headpic{                                           % UM heading
        \indent
        \setlength{\unitlength}{.4mm}
        \thinlines
        \par
        \begin{picture}(29,16)
        \put(165,16){\line(1,0){4}}
        \put(170,16){\line(1,0){4}}
        \put(180,16){\line(1,0){4}}
        \put(175,0){\line(1,0){4}}
        \put(180,0){\line(1,0){4}}
        \put(185,0){\line(1,0){4}}
        \put(169,0){\line(0,1){16}}
        \put(170,0){\line(0,1){16}}
        \put(179,0){\line(0,1){16}}
        \put(180,0){\line(0,1){16}}
        \put(184,0){\line(0,1){16}}
        \put(185,0){\line(0,1){16}}
        \put(169,16){\oval(8,32)[bl]}
        \put(170,16){\oval(8,32)[br]}
        \put(179,0){\oval(8,32)[tl]}
        \put(185,0){\oval(8,32)[tr]}
        \end{picture}
        \par\vskip-6.5mm
        \thicklines}
\def\title#1#2#3#4{\border\headpic {\hbox to\hsize{#4 \hfill UMDEPP #3}}\par
        \begin{center} \vglue .5in {\large\bf #1}\\[.6in]
        {#2}\\[.1in] {\it Department of Physics and Astronomy}\\
        {\it University of Maryland, College Park, MD 20742}\\[1.5in]
        {\bf ABSTRACT}\\[.1in] \end{center} \begin{quotation}}  % title stuff
\def\Title#1#2#3#4#5#6#7{\border\headpic
        {\hbox to\hsize{#7 \hfill UMDEPP #6}}\par
        \begin{center} \vglue .4in {\large\bf #1}\\[.4in]
        {#2}\\[.1in] {\it Department of Physics and Astronomy}\\
        {\it University of Maryland, College Park, MD 20742}\\[.1in]
        {#3}\\[.1in] {\it {#4}}\\ {\it {#5}}\\[.5in] {\bf ABSTRACT}\\[.1in]
        \end{center} \begin{quotation}}                 % " for 2 authors
\def\endtitle{\end{quotation}\newpage}                  % end title page

\def\dpx{{{ =\hskip-3.75pt{\lin}}\hskip3.75pt }}

\def\qd{{\kern0.5pt
                   q \kern-5.05pt \raise5.8pt\hbox{$\textstyle.$}\kern
0.5pt}}

\begin{document}

\def\gfrac#1#2{\frac {\scriptstyle{#1}}
        {\mbox{\raisebox{-.6ex}{$\scriptstyle{#2}$}}}}
\def\gg{{\hbox{\sc g}}}
\border\headpic {\hbox to\hsize{August 1995 \hfill {UMDEPP 96-19}}}
\par
\setlength{\oddsidemargin}{0.3in}
\setlength{\evensidemargin}{-0.3in}
\begin{center}
\vglue .08in
{\large\bf Why Auxiliary Fields Matter: The Strange\\
Case of the 4D, N = 1 Supersymmetric\\
QCD Effective Action\footnote {Supported in part by National
Science Foundation Grant PHY-91-19746 \newline ${~~~~~}$ and by NATO
Grant CRG-93-0789}  }
\\[.72in]

S. James Gates, Jr.
\\[.02in]
{\it Department of Physics\\
University of Maryland\\
College Park, MD 20742-4111  USA}\\[.2in]
{\bf {\tt gates@umdhep.umd.edu}}\\[2in]

{\bf ABSTRACT}\\[.002in]
\end{center}
\begin{quotation}
{Within a four dimensional manifestly N = 1 supersymmetric action, we show
that Wess-Zumino-Novikov-Witten (WZNW) terms can be embedded in an
extraordinarily simple manner into a purely chiral superaction. In order to
achieve this result it is necessary to assign spin-0 and spin-1/2 degrees
of freedom both to chiral superfields and as well to non-minimal scalar
multiplets.  We propose a new formulation for the effective low-energy
action of 4D, N = 1 supersymmetric QCD that is consistent with holomorphy
through fourth order in the pion superfield. After reduction to a 2D, N = 2
theory we find a new class of manifestly supersymmetric non-linear
$\s$-models with torsion.}

\endtitle
%%%%%%%%%%%%%%%%%%%%%%%%%%%%%%%%%%%%%%%%%%%%%%%%%%%%%%%%%%%%%%%%%%
\section{Introduction}
%%%%%%%%%%%%%%%%%%%%%%%%%%%%%%%%%%%%%%%%%%%%%%%%%%%%%%%%%%%%%%%%%%%%

{}~~~~Over a decade ago \cite{A}, there began efforts to utilize supersymmetric
models to construct the successor to the standard model.  These efforts
received
a further boost with the realization \cite{B} that such theories seem naturally
to
occur as the low-energy limit of four dimensional superstring and heterotic
string theories. A brief survey of the literature would lead one to believe
that there are no unresolved issues in how 4D, N = 1 superfields occur in
this limit. In fact, there are a number of {\it {assumptions}} that are most
often {\underline {not}} {\underline {even}} stated in presentations of the
low-energy action (purportedly {\it {derived}} from superstrings) upon
which most model building is based and phenomenology elucidated.  One of
these assumptions is that the spin-0 and spin-1/2 fields that are derived
from the spectrum of string theory necessarily are described by 4D, N = 1
chiral superfields.  It is not generally recognized that this is just an
assumption. The reason why this {\underline {is}} an assumption lies in the
fact that there exist little recognized alternative 4D, N = 1 superfields,
the non-minimal scalar multiplet \cite{C} being one, that contain exactly
the same on-shell spectrum as the usual chiral multiplet. We named such
off-shell representations of 4D, N = 1 supersymmetry ``variant
representations.''
Although the non-minimal multiplet has exactly the same on-shell spectrum as
the chiral multiplet, it contains a very different set of auxiliary fields.
As we pointed out previously, the non-minimal scalar multiplet can appear
as an alternate to the usual N = 1 K\" ahler non-linear $\s$-models and
as well interact with the usual chiral multiplets \cite{D}.  Among these latter
interactions there is a curious result that if a non-minimal scalar multiplet
gains a mass, it can {\underline {only}} do so in tandem with a chiral
multiplet!
In other words, this mechanism provides a natural explanation for the
occurrence
of Dirac spinors within the context of 4D, N = 1 supersymmetric models.

In most discussions of supersymmetric theories, the issue of auxiliary
fields is treated in a cavalier fashion.  One would think that there is no
essentially important role played by auxiliary fields.
Nothing could be further from the truth. One reason this attitude prevails
is that there have been few demonstrations of just what dynamical consequences
exist when the off-shell spectrum of two multiplets with the same on-shell
spectrum are compared. A place where such differences can be shown to have
demonstrable consequences is non-linear $\s$-models. Similarly differences
can also be observed in higher derivative theories.  Typically, what occurs
is that the fields that are usually considered ``auxiliary'' can become
propagating.  Under this circumstance, clearly the structure of the auxiliary
fields is important. Higher derivative theories are typically characteristic
of effective field theories.  Among these, perhaps the most important is the
low-energy effective Lagrangian ${\cal L}_{eff}$ of QCD.  It is known that
leading terms of this theory are described by a chiral SU(3) $\otimes$ SU(3)
non-linear $\s$-model.   Another term of ${\cal L}_{eff}$ is the
Wess-Zumino-Novikov-Witten term (WZNW) \cite{E}.

Along these lines there has been some discussion of what is the structure of
the 4D, N = 1 supersymmetric extension of the WZNW term \cite{F}.  It is the
purpose of this note to show that the introduction of non-minimal scalar
multiplets, to describe some of the spin-0 and spin-1/2 fields in the
effective action, opens an alternate formulation of the 4D, N = 1
supersymmetric
WZNW term. This result highlights the importance of auxiliary fields. Our
result
also provides the most striking evidence to date that the assumption that
{\underline {only}} chiral superfields describe the matter seen in Nature is
incorrect.

%%%%%%%%%%%%%%%%%%%%%%%%%%%%%%%%%%%%%%%%%%%%%%%%%%%%%%%%%%%%%%%%%%%%
\section{Chiral and Non-minimal Multiplet WZNW \newline
 Theory}
%%%%%%%%%%%%%%%%%%%%%%%%%%%%%%%%%%%%%%%%%%%%%%%%%%%%%%%%%%%%%%%%%%%

{}~~~~Almost every researcher who has investigated four dimensional
N = 1 supersymmetry is aware of the chiral scalar or Wess-Zumino multiplet
\cite{G}.  The multiplet is described by a chiral superfield $\Phi$ ($
{\Bar D}_{\dot \a} \Phi = 0$). The component fields are defined by (we use
{\it {Superspace}} conventions \cite{D})
$$
A ~\equiv~  \Phi \, | ~~~~,~~~~ \psi_{\a} ~\equiv~ {D}_{\a} \Phi \, | ~~~~,~~~~
F ~\equiv~ {D}^2 \Phi \, | ~~~~,
\eqno(2.1) $$
and appear in the usual linear action as
$$
{\cal S}_{WZ} ~=~ \int d^4 x d^2 \q d^2 {\Bar \q} ~ {\Bar \Phi} \, \Phi ~=~
\int d^4 x ~ \Big[ \, - (\pa^{\underline a} {\Bar A} \, ) (
\pa_{\underline a} A \, ) ~-~ i {\Bar \psi}^{\dot \a} \pa_{\underline a}
{\psi}^{\a} ~+~ {\Bar F} F \, \Big]  ~~~~.
\eqno(2.2) $$

The non-minimal scalar multiplet is described by a complex linear superfield
$\S$ (subject to the constraint $ {\Bar D}^2 \S = 0$). The component fields are
defined by
$$ \eqalign{
B ~\equiv~  \S \, | ~~~&,~~~~ {\Bar \zeta}_{\dot \a} ~\equiv~ {\Bar D}_{\dot
\a} \S \, | ~~~~, \cr
{\r}_{\a} ~\equiv~ {D}_{\a}  \S \, | ~~~&,~~~~ H ~\equiv~
{D}^2 \S \, | ~~~~,\cr
p_{\underline a} ~\equiv~ {\Bar D}_{\dot \a} D_{\a} \S \, | ~~~&,~~~~
{\Bar \b}_{\dot \a} ~\equiv~ \frac 12  D^{\a} {\Bar D}_{\dot \a} D_{\a} \S
\, | ~~~~, }
\eqno(2.3) $$
and appear in the usual linear action as
$$ \eqalign{
{\cal S}_{NM} &=~ -~ \int d^4 x d^2 \q d^2 {\Bar \q} ~ {\Bar \S} \, \S \cr
&=~ \int d^4 x ~ \Big[ \, - (\pa^{\underline a} {\Bar B} \, ) (
\pa_{\underline a} B \, ) ~-~ i {\Bar \zeta}^{\dot \a} \pa_{\underline a}
{\zeta}^{\a} ~-~ {\Bar H} H \cr
&{~~~~~~~~~~~~~~~}\, ~+~ 2 \, {\Bar p}^{\underline a} p_{\underline a}  ~+~
{\b}^{\a} {\r}_{ \a} ~+~ {\Bar \b}^{\dot \a} {\Bar \r}_{\dot \a}
\, \Big]  ~~~~. }
\eqno(2.4) $$
As is apparent from the last result above, only $B$ and ${\zeta}^{\a}$ are the
propagating fields among the off-shell 12 + 12 (bosons + fermions) degrees of
freedom of the non-minimal scalar multiplet.

At this point we recall for the reader results in 2D, N = 2 superfield theory
\cite{G1}. Within this class of theories, it is known that there are two
{\it {distinct}} minimal scalar multiplets, chiral multiplets and twisted
chiral multiplets \cite{H}.  The superfield form of the linear kinetic term
for the twisted chiral multiplet has a minus sign in comparison to that of
the chiral multiplet. We see exactly the same behavior above for the 4D
chiral and non-minimal superfield actions.   In a 2D, N = 2 non-linear
$\s$-model theory with manifest supersymmetry, the only known way to
introduce torsion requires the simultaneous presence of both chiral and
twisted chiral superfields. In 4D, the analog of the 2D torsion term is
provided by the WZNW term. Thus, it is natural to suggest that we should
be able to introduce a 4D, N = 1 supersymmetric WZNW term by utilizing
both chiral and non-minimal multiplets.

The starting point in the implementation of this proposal is to note that
the condition that $\S$ is a complex linear superfield (i.e. ${\Bar D}^2 \S
= 0$) necessarily implies that the quantity ${\Bar D}_{\dot \a} \S $ is a
chiral superfield and can therefore lead to a supersymmetric invariant in
an F-term!  So we introduce a number of chiral superfields $\Phi^{\rm I}$
along with an equal number of non-minimal scalar superfields $\S^{\rm I}$
where ${\rm I} = 1,..., m$. We also require the existence of a fourth order
tensor that is a function of the chiral superfields.  We denote this tensor
by ${\cal J}_{\rm I \, J \, K \, L} (\Phi)$.  It follows that the term
below is a supersymmetric invariant
$$
{\cal S}_{WZNW} ~=~  \frac 14 \,  \int d^4 x \, d^2 \q ~ {\cal J}_{\rm I \,
J \, K \, L} (\Phi) ({\Bar D}^{\dot \a} \S^{\rm I} \, ) \, ({\Bar D}^{\dot
\b} \S^{\rm J} \, ) \, (\pa^{\g} {}_{\dot \a} \Phi^{\rm K} \,) \, ( \pa_{\g
\dot \b} \Phi^{\rm L} \,)  ~+~ {\rm h.}{\rm c.} ~~~~.
\eqno(2.5) $$
Let us note that the most general non-linear $\s$-model term involving these
superfields takes the form,
$$
{\cal S}_{\s} ~=~  \int d^4 x d^2 \q d^2 {\Bar \q} ~ {\Hat \O} (\, \Phi, {\Bar
\Phi} ;  \S, {\Bar \S} \, ) ~~~~.
\eqno(2.6) $$
The function ${\Hat \O}$ is similar to a K\" ahler potential.  However, as
shown in the the latter work of \cite{D}, the metric for the space for which
$\S^{\rm I}$ provides coordinates is not of the form of a K\" ahler metric.
In fact, the metric for the $\S^{\rm I}$-space is {\underline {not}} even of
hermitian form in general.  We thus have a counter-example to the well known
folklore that 4D, N = 1 supersymmetric non-linear $\s$-models necessarily
describe K\" ahler manifolds.  (The global description of the complex space
described solely by $\S$-coordinates has never been given.)  Note that one
special choice\footnote{The choice of this function is ultimately done to
produce the best phenomenological fit.} of the function ${\Hat \O}$ is given
by a fibration in which the $\S$-coordinates are fibers over a space with
the $\Phi$-coordinates. Such a space is described by
$$
{\Hat \O} ~=~ \frac 12 \, [ \, g_{\rm {I\, J}}(\Phi) ~+~ {\Bar g}_{\rm {J\, I}}
({\Bar \Phi}) \,] \, [~ {\Bar \Phi}^{\rm I} {\Phi}^{\rm J} ~-~
{\Bar \S}^{\rm I} {\S}^{\rm J}  ~] ~~~~,
\eqno(2.7) $$
in terms of a holomorphic function $g_{\rm {I\, J}}(\Phi)$ (for one choice
of this function see appendix A). In the limit where we set the non-minimal
multiplets to zero, we see that the chiral superfields have a special
K\" ahler geometry\footnote{The first appearance of this type of geometry
is in \cite{G1}.}. The limiting K\" ahler potential $K(\Phi, {\Bar \Phi})$
can be written in the form $K(\Phi, {\Bar \Phi}) = \frac 12 [ {\Bar \Phi}^{
\rm I} g_{\rm {I\, J}}(\Phi) {\Phi}^{\rm J} + {\Phi}^{\rm J} {\Bar g}_{\rm
{I\, J}} ({\Bar \Phi}) {\Bar \Phi}^{\rm I}] $ in order to make the special
K\" ahler geometry for the chiral superfields manifest. Defining ${\Tilde
\Phi}_{\rm I} \equiv  g_{\rm {I\, J}}(\Phi)  {\Phi}^{\rm J}$, we can re-write
$K(\Phi,
{\Bar \Phi})$ in the form $K(\Phi, {\Bar \Phi}) = \frac 12 [{\Bar \Phi}^{\rm
I} {\Tilde \Phi}_{\rm I} + {\Phi}^{\rm I} {\Tilde {\Bar \Phi}}_{\rm I}]$ in
order to make contact with the recent work on exact results for N = 2
supersymmetric Yang-Mills effective actions \cite{I}. This form also makes
obvious the presence of the duality pairs $({\Phi}^{\rm I} , \, {\Tilde
\Phi}_{\rm I})$ that are related by elliptic curves.

Let us offer another interpretation of (2.5), (2.7) and (2.8) below. Within
the confines of 2D, N = 2 superconformal field theory, there have been found
to exist (c,c) rings and (a,c) rings. The former correspond to functions of
chiral multiplets while the latter correspond to twisted chiral multiplets.
The interesting point is that (a,c) rings were discovered much later than
(c,c) rings. This discovery of these distinct supersymmetry representations
in the spectrum of the theory occurred even though they were not put in as
elementary representations on the 2D world sheet. This example shows
us that non-perturbatively supersymmetric systems are able to generate
states that are distinct supersymmetry representations from the elementary
states. On the other hand, if (a,c) rings are included at the elementary level,
then 2D, N = 2 superconformal field theories can possess an additional
symmetry, i.e. mirror symmetry.  This suggests that the non-minimal scalar
multiplet may be generated non-perturbatively in 4D and if they are included
in the underlying supersymmetric renormalizable QCD theory, it may possess a
larger symmetry group.

Thus, we should be able to embed the QCD low-energy effective
action into a supersymmetric action of the form
$$
{\cal S}_{eff} ~=~ {\cal S}_{\s} ~+~ {\cal S}_{WZNW} ~~~~.
\eqno(2.8) $$
In the next section we will look at the component formulation that follows
from the proposal above. However, in closing this section, we note that our
proposed description of the 4D, N = 1 supersymmetric QCD low-energy effective
action with WZNW term is the {\underline {first}} that is consistent with
holomorphy \cite{I}, the concept that holomorphic functions determine
the effective action. In fact, we gave the {\underline {first}} demonstration
\cite{G1} that the 4D, N = 2 supersymmetric Yang-Mills action is classically
determined by holomorphic functions.  Recently, major advances have occurred
in understanding the 4D, N = 2 supersymmetric Yang-Mills effective action
due to the presence of holomorphy and proposals have been made that it
should play a role in increasing our understanding of the 4D, N = 1
supersymmetric Yang-Mills effective action.

%%%%%%%%%%%%%%%%%%%%%%%%%%%%%%%%%%%%%%%%%%%%%%%%%%%%%%%%%%%%%%%%%%%%
\section{Embedding ${\cal L}_{eff} (QCD) $ in a 4D, N = 1 \newline
${}$ Supersymmetric Theory}
%%%%%%%%%%%%%%%%%%%%%%%%%%%%%%%%%%%%%%%%%%%%%%%%%%%%%%%%%%%%%%%%%%%

{}~~~~The calculation of the component results follows using the by now
well established projection technique. We find ${\cal S}_{WZNW}$ leads to
$$ \eqalign{
 &   \frac 14 \, \int d^4 x d^2 \q ~ {\cal J}_{\rm I \, J \, K \, L} (
\Phi) ({\Bar D}^{\dot \a} \S^{\rm I} \, ) ({\Bar D}^{\dot \b} \S^{\rm J} \,
) (\pa^{\g}{}_{\dot \a} \Phi^{\rm K} \,) ( \pa_{\g \dot \b} \Phi^{\rm L} \,)
{~~~~~~~~~~}  \cr
 &=~   \frac 14 \,  \int d^4 x \Big[ \, - {\cal J}_{\rm I \, J \, K \, L} (A)
\, ( i 2 \pa^{\a \dot \a} B^{\rm I} ~-~ p^{\a \dot \a ~ \rm I} \, )
( i 2 \pa_{\a}{}^{\dot \b} B^{\rm J} ~-~ p_{\a} {}^{\dot \b ~ \rm J} \, )
(\pa^{\g}{}_{\dot \a} A^{\rm K} \,) ( \pa_{\g \dot \b} A^{\rm L} \,) \cr
& {~~~~~~~~~~~~~~~} +~  2 {\cal J}_{\rm I \, J \, K \, L} (A)\, ( i \pa_{\a}
{}^{\dot \a} \r^{\a \, {\rm I}} ~-~ \b^{\dot \a \, {\rm I}} \,)
\,{\Bar \zeta}^{\dot \b \, {\rm J}} (\pa^{\g}{}_{\dot \a} A^{\rm K} \,)
 ( \pa_{\g \dot \b} A^{\rm L} \,) \cr
& {~~~~~~~~~~~~~~~} +~  {\cal J}_{\rm I \, J \, K \, L} (A) \,{\Bar
\zeta}^{\dot
\a \, {\rm I}}{\Bar \zeta}^{\dot \b \, {\rm J}} [\, (\pa^{\g}{}_{\dot \a}
\psi^{\a
\rm K} \,) ( \pa_{\g \dot \b} \psi_{\a}{}^{\rm L} \,) ~+~ 2
(\pa^{\g}{}_{\dot \a} A^{\rm K} \,) ( \pa_{\g \dot \b} F^{\rm L}
\,) \,] \cr
& {~~~~~~~~~~~~~~~} +~  4 {\cal J}_{\rm I \, J \, K \, L} (A) \,
( i 2 \pa^{\a \dot \a} B^{\rm I} ~-~ p^{\a \dot \a ~ \rm I} \, ) \,
{\Bar \zeta}^{\dot \b \, {\rm J}} (\pa^{\g}{}_{\dot \a} \psi_{\a}^{
\rm K} \,)  ( \pa_{\g \dot \b} A^{\rm L} \,) \cr
& {~~~~~~~~~~~~~~~} +~  2 {\cal J}_{\rm I \, J \, K \, L \, , {\rm M}} (A) \,
\psi^{\a \rm M} \, {\Bar \zeta}^{\dot  \a \, {\rm I}}{\Bar \zeta}^{\dot \b \,
{\rm J}} (\pa^{\g}{}_{\dot \a} \psi_{\a}^{\rm K} \,)  ( \pa_{\g \dot \b}
A^{\rm L} \,) \cr
& {~~~~~~~~~~~~~~~} -~  2 {\cal J}_{\rm I \, J \, K \, L \, , {\rm M}} (A) \,
\psi^{\a \rm M} ( i 2 \pa^{\a \dot \a} B^{\rm I} ~-~ p^{\a \dot \a ~ \rm I}
\, ) \, {\Bar \zeta}^{\dot \b \, {\rm J}} (\pa^{\g}{}_{\dot \a} A^{\rm K} \,)
( \pa_{\g \dot \b} A^{\rm L} \,) \cr
& {~~~~~~~~~~~~~~~} +~ {\cal J}_{\rm I \, J \, K \, L \, , {\rm M}} (A) \,
F^{\rm M} \,{\Bar \zeta}^{\dot  \a \, {\rm I}}{\Bar \zeta}^{\dot \b \,
{\rm J}} (\pa^{\g}{}_{\dot \a} A^{\rm K} \,) ( \pa_{\g \dot \b} A^{\rm L}
\,) ~ \Big]
{}~~~~. }
\eqno(3.1) $$
As can be seen, only the first line of the rhs consists of purely bosonic
terms.  Let us focus our analysis by only considering these terms.

It is our first observation that if we set the auxiliary field $p_{
\underline a}$ to zero, then the purely bosonic terms collapse to
$$ \eqalign{   \frac 14 \,
\int d^4 x \, d^2 \q ~ &{\cal J}_{\rm I \, J \, K \, L} (\Phi) \,
({\Bar D}^{\dot \a} \S^{\rm I} \, ) \, ({\Bar D}^{\dot \b} \S^{\rm J} \, ) \,
(\pa^{\g} {}_{\dot \a} \Phi^{\rm K} \,) \, ( \pa_{\g \dot \b} \Phi^{\rm L}
\,) |_{phys.\, fields}{~~~~~~~~~~}  \cr
 &=~  \int d^4 x \Big[ \,  {\cal J}_{\rm I \, J \, K \, L} (A) \, (
\pa^{\a \dot \a} B^{\rm I} \, ) \, ( \pa_{\a}{}^{\dot \b} B^{\rm J} \, ) \,
(\pa^{\g}{}_{\dot \a} A^{\rm K} \,) \, ( \pa_{\g \dot \b} A^{\rm L} \,) \,
\Big]  \cr
 &=~  \int d^4 x \Big[ \,  {\cal J}_{\rm I \, J \, K \, L} (A) \,
{\rm P}^{\underline a \underline b \underline c  \underline d} \,
( \pa_{\underline a} B^{\rm I} \, ) \, ( \pa_{\underline b} B^{\rm J} \, )
\, (\pa_{\underline c} A^{\rm K} \,) \, ( \pa_{\underline d} A^{\rm L} \,)
\, \Big]  ~~~~. }
\eqno(3.2) $$
where ${\rm P}^{\underline a \underline b \underline c  \underline d}
\equiv [ \eta^{\underline a [ \underline c}  \eta^{\underline d ]
\underline b} ~+~ i \e^{\underline a \underline b \underline c
\underline d} ] $.  Up until this point, we have not made any assumption
regarding the explicit form of ${\cal J}_{\rm I \, J \, K \, L} (A)$.  We
could easily choose it to be the (4,0) form that is defined in the
non-supersymmetric component WZNW action (see appendix A). However, (3.2)
has the consequence that it can describe both the WZNW term as well as
the Skyrme term. In the following we we simply concentrate on the WZNW
term and thus we choose ${\cal J}_{\rm I \, J \, K \, L} (A)$ to be
define by (A.6)\footnote{A few minor modifications are required in the
supersymmetric case.}. Since $p_{\underline a}$ actually has a more
complicated equation of motion that depends on the leading term of the
effective action, its elimination will produce other higher order
interactions. However, their presence does not disturb our present results.
These and a number of other details will be discussed in a future work.

Now we want the component pion fields that are contained in our QCD
superfield WZNW term of (2.8) to agree precisely the non-supersymmetric
QCD effective action (see (A.7)). This will be the case if the following
identifications are made,
$$\Phi | ~=~ {\cal A} (x) ~+~ i \, [ \, \Pi (x) ~+~ \Theta (x) \,] ~~~,~~~
\S | ~=~ {\cal B} (x) ~+~ i \, [ \, \Pi (x) ~-~ \Theta (x) \,] ~~~.
\eqno(3.3) $$
where $\Pi (x)$ is the pion octet. Thus, we see that the pion superfield
is a linear mixture of chiral and complex linear superfields.
This is analogous to the fact that a Dirac field in a supersymmetric theory
can only occur as a linear combination of basic superfields. We are thus
motivated to define the super-pion superfield by
$$ \Pi ~\equiv~ -i  \frac 14 \Big[ \, \Phi ~+~ \S  ~-~
{\Bar  \Phi} ~-~ {\Bar \S} \, \Big]  ~~~~.
\eqno(3.4) $$
By the same token we see that in a manifestly supersymmetric world, in
addition to the super-pion, there are mirror super-pions defined by
$$ \Theta ~\equiv~ -i  \frac 14 \Big[ \, \Phi ~-~ \S  ~-~
{\Bar  \Phi} ~+~ {\Bar \S} \, \Big]   ~~~~.
\eqno(3.5) $$
There are also parity doubles of these fields that are most conveniently
defined by
$$ {\cal A} ~\equiv~ \frac 12 \Big[ \, \Phi ~+~ {\Bar  \Phi} \, \Big]  ~~~~,
{}~~~~ {\cal B} ~\equiv~ \frac 12 \Big[ \, \S ~+~ {\Bar \S}  \, \Big]  ~~~~.
 \eqno(3.6) $$
Similarly, applying various spinor derivatives to these superfields
produce the spin-1/2 pionino SU(3) multiplet and their parity doubles.
Here we have some ambiquity. We have enough spinor components to form
a Dirac pionino SU(3) multiplet or two Majorana pionino SU(3) multiplets.
In the former case, the pionini are isomorphic to the baryon octet that
contains the proton!

The leading term in (2.8) will also contain exactly the leading term
of (A.7) if we identify the function $g_{\rm I \, J}$\footnote{Once again a
few minor modifications are required in the supersymmetric case.} that appears
in (2.7) with that defined in (A.4). Thus, we find that there is a very simple
embedding of ${\cal L}_{eff} (QCD)$ into our superfield theory.

%%%%%%%%%%%%%%%%%%%%%%%%%%%%%%%%%%%%%%%%%%%%%%%%%%%%%%%%%%%%%%%%%%%%
\section{ Conclusion}
%%%%%%%%%%%%%%%%%%%%%%%%%%%%%%%%%%%%%%%%%%%%%%%%%%%%%%%%%%%%%%%%%%%%

{}~~~~At this point, it is useful to compare our new suggestion for a 4D,
N = 1 supersymmetric extension of the WZNW terms to those that exist in
the prior literature. The relevant work occurred in reference \cite{F}.
There it was proposed that the 4D, N = 1 supersymmetric extension of the
WZNW term is of the form
$${\cal S}_{WZNW} ~=~  \int d^4 x d^2 \q  d^2 {\Bar \q} ~ \Big[ \, \b_{\rm
{ I \,J  {\bar {\rm K}}}} ( D^{\a} \Phi^{\rm I}\,) ( \pa_{\a \dot \b}  \Phi^{
\rm J} \,) ( {\Bar D}^{\dot \b} {\Bar \Phi}^{\bar {\rm K}} \,) ~+~ {\rm h.}
{\rm c.}  ~\Big] ~~~~.
\eqno(4.1) $$
If we compare our results to the older ones, several features are apparent.
Foremost, the previous result utilizes an action that is integrated over the
full superspace. (This means for example that all of the chiral superfields
contained in (4.1) could be replaced by complex linear superfields and we
would then obtain another WZNW-type term.)  In particular, the quantity
$\b_{\rm { I \,J  {\bar {\rm K}}}}$ is {\underline {not}} holomorphic.
Our choice need only be integrated over a chiral subspace due to its chirality
(i.e. holomorphicity).  At the level of component fields, the differences
are simply tremendous! Our suggestion contains many fewer terms. At most four
fermion but not six fermion terms appear in our construction in contrast to
(4.1). Finally, there are terms in (4.1) that are quartic in temporal
derivatives of bosonic fields. In our proposal no such terms of this high
order in temporal derivatives appear.  This last point is rather telling.
It is certainly true that the non-supersymmetric WZNW terms contains no more
than first order temporal derivatives.

Ordinary 4D, N = 1 chiral and non-minimal multiplets possess an uncanny
resemblance to 2D, N = 2 chiral and twisted chiral multiplets. This
naturally raises the question of whether there might exist some 4D, N = 1
analog to mirror symmetry. We could formally define a 4D mirror operator
that sends chiral multiplets into non-minimal multiplets and vice-versa.
There are important differences, however. Off-shell chiral and non-minimal
multiplets do not possess the same number of degrees of freedom. So there
are some issues that require additional study.  Finally, we believe that
our result regarding the simple embedding of ${\cal L}_{eff} (QCD)$ should
act as a warning that the sole use of chiral multiplets to describe
matter is not always wise. We re-emphasize the cautionary note we made
along these lines previously in the second work of
\cite{D}.

The problem our presentation demonstrates has its ultimate cause
in our lack of mastery of string theory.  As
presently formulated, we simply do not possess a direct (i.e.
without making any assumptions) way to derive from string theory the
off-shell superfields that presumably emerge in its low-energy
limit. For some time, we have believed that it is quite likely
that non-minimal scalar multiplets must be involved in this limit.
Our reason for this belief is that it appears likely that the
4D, N = 2 low-energy limit of string theory contains at least some
non-minimal scalar multiplets! The only known off-shell formulation
of 4D, N = 2 hypermultiplets \cite{ST} contains 4D, N = 1
non-minimal scalar multiplets.  Finally, it is interesting to ponder
further WZNW extensions to 4D, N = 2
supersymmetry. The recent advances \cite{I} are silent on
the 4D, N = 2 WZNW term. Here we would like to know
if the {\it {two}} distinct 4D, N = 2 hypermultiplets (\cite{ST} and
\cite{FY}) play roles analogous to that of the 4D, N = 1 chiral
and non-minmal multiplets in the 4D, N = 1 WZNW term.

This latest result together with the ``natural Dirac mass'' associated
with a pairing of a chiral superfield together with a complex linear
superfield (i.e. $(\Phi^{\rm I} , \, \S^{\rm I} \, )$) seems to be
hinting that there is something truly fundamental but not understood
occurring.  As we noted previously, the current generation of
supersymmetric phenomenological models totally ignores the possibility
that ordinary matter may contain such pairings.   We can well imagine
scenarios in which one chiral
part of a Dirac particle is assigned to chiral superfields and the
other chiral part of the same Dirac particle is assigned to complex
linear superfields.  This might well serve as an intrinsic reason
why chiral asymmetry occurs in supersymmetric extensions of the
standard model and as well could easily provide the long sought
use of supersymmetry to protect the vanishing masses of neutrini.
Indeed, if supersymmetry is observed in Nature this could make an
attractive explanation for why handedness matters in our universe!
$$\eqalign{ ~~ &{~} \cr &{~} \cr  &{~} \cr &{~} \cr
 &{~} \cr &{~} \cr  &{~} \cr &{~} \cr   &{~} \cr &{~} \cr } $$

\noindent
%%%%%%%%%%%%%%%%%%%%%%%%%%%%%%%%%%%%%%%%%%%%%%%%%%%%%%%%%%%%%
{\bf {Acknowledgment; }} \newline \noindent
%%%%%%%%%%%%%%%%%%%%%%%%%%%%%%%%%%%%%%%%%%%%%%%%%%%%%%%%%%%%%
I wish to thank Ms. Lubna Rana for useful discussions.

\newpage \noindent
%%%%%%%%%%%%%%%%%%%%%%%%%%%%%%%%%%%%%%%%%%%%%%%%%%%%%%%%%%%%%
{\bf {Appendix A:  Brief Review of ${\cal L}_{eff}(QCD)$ }}
%%%%%%%%%%%%%%%%%%%%%%%%%%%%%%%%%%%%%%%%%%%%%%%%%%%%%%%%%%%%%

In this very brief appendix we simply gather together the basic
facts concerning the low-energy effective action for QCD. We begin
with a definition of the SU(3) pion octet
$$ \frac 1{f_{\pi}} \Pi ~\equiv~
\frac 1{ f_{\pi}} \Pi^i \l_i ~=~ \frac 1{ f_{\pi}} \left(\begin{array}{ccc}
{}~\frac{\pi^0}{\sqrt 2} ~+~ \frac{\eta}{\sqrt 6} & ~~\pi^+ &  ~~K^+ \\
{}~\pi^- & ~~-\, \frac{\pi^0}{\sqrt 2} ~+~ \frac{\eta}{\sqrt 6} &  ~~K^0\\
{}~K^- & ~~{\Bar K}^0 &  ~~ - \eta \sqrt {\frac 23}     \\
\end{array}\right) ~~~~.
\eqno(A.1) $$
Here $\l_1 , \,..., \l_8$ are the Gell-Mann SU(3) matrices. Further $f_{\pi
}$ is the weak pion coupling constant\footnote{It should be noted that we
differ from Witten's convention of this parameter by a factor of two.}
with the dimensions of mass. Group elements are formed by writing $U(\Pi)
= \exp [\, i {f_{\pi}}^{-1}\Pi \,]$.  We define left ($ L_m {}^i (\Pi)$)
and right ($ R_m {}^i (\Pi)$) Maurer-Cartan forms by the equations
$$
U^{-1} \pa_{\underline a} U ~=~ i  {f_{\pi}}^{-1} ( \, \pa_{\underline a} \Pi^m
\,)~ L_m {}^i (\Pi) \, \l_i  ~~~~,~~~~ (\, \pa_{\underline a} U \,)  U^{-1} ~=~
i {f_{\pi}}^{-1} ( \, \pa_{\underline a} \Pi^m  \,) ~R_m {}^i
(\Pi) \, \l_i  ~~~~.
\eqno(A.2) $$
These definitions allow $L_m {}^i (\Pi)$ and $R_m {}^i (\Pi)$ to be calculated
as power series in $\Pi^i$ from \cite{ACG}
$$\eqalign{
L_m {}^i (\Pi) &\equiv ~ (C_2)^{-1} {\rm {Tr}} \Big[\,  T^i \Big(
\frac { 1 ~-~ e^{-\D} }{\D} \Big) T_m \Big] ~~~, \cr
R_m {}^i (\Pi) &\equiv ~ (C_2)^{-1} {\rm {Tr}} \Big[\,  T^i \Big(
\frac { e^{\D} ~-~ 1 }{\D} \Big) T_m \Big] ~~~, }
\eqno(A.3) $$
where $\D T_m \equiv  i {f_{\pi}}^{-1} [ \Pi \, , \, T_m  ]$, $\D^2 T_m = \D \D
T_m$, etc. and the constant $C_2$ is determined so that $ L_m {}^i (0) = R_m
{}^i
(0) = \d_m {}^i$.   As a consequence we see
$$ {\cal S}_{\s} ~=~   \frac {f_{\pi}^2}{2 C_2} \, \int d^4 x ~
{\rm {Tr}} [\, ( \pa^{\underline a} U \,) ~ (\pa_{\underline a} U ^{-1} \,) \,]
{}~=~ \frac 12 \, \int d^4 x ~ \, g_{m \, n} \, (\Pi) ~ ( \pa^{\underline
a} \Pi^m \,) ~ (\pa_{\underline a} \Pi^n \,) ~~~,
\eqno(A.4) $$
where $g_{m \, n}  = \d_{i \, j} \, L_m {}^i \, L_n {}^j = \d_{i \, j} \, R_m
{}^i \, R_n {}^j$.

The remaining well known term in the QCD effective action is described by the
WZNW term. We follow Witten \cite{E} who, using the Vainberg technique
\cite{V},
showed that with an appropriate normalization this term possesses an
integer quantized coefficient, $N_C$.  Using a  real parameter $y$ that
takes on values between 0 and 1, we define an extended group element $\Hat U$
through the relation $\Hat U = \exp [\, i y f_{\pi}^{-1} \Pi \,]$.
In terms of the extended group element, the WZNW term is given by
$$ \eqalign{ {~~~~~~~~}
{\cal S}_{WZNW} &=~ - i N_C \, [ \, 2 {~}_{\dot {~}} 5! \, ]^{-1}
\int d^4 x \, \int_0^1 d y ~ {\rm {Tr}} \Big[ \, ( {\Hat U}^{-1} \pa_y {\Hat U}
\,) ~ {\Hat {\cal W}}_4 \, \Big]   ~~~~, \cr
{\Hat {\cal W}}_4 &=~ \e^{{\underline a}{\underline b}{\underline c}{\underline
d}} \, (\pa_{\underline a} {\Hat U} ^{-1} \,) \, (\pa_{\underline b} {\Hat U}
\,) \, (\pa_{\underline c} {\Hat U}^{-1} \,) \,  (\pa_{\underline d} {\Hat U}
\,) \,   ~~~~. }  \eqno(A.5) $$
or more directly using the elements of the pion octet this just becomes
$$ \eqalign{ {~~~~}
{\cal S}_{WZNW} &=~ \int d^4 x \, \e^{{\underline a}{\underline b}{\underline
c}
{\underline d}}{\cal J}_{m \, n \, r \, s} (\Pi)
(\pa_{\underline a} \Pi^m \,) \, (\pa_{\underline b} \Pi^n \,) \,
(\pa_{\underline c} \Pi^r \,)  \, (\pa_{\underline d} \Pi^s \,) ~~~~, \cr
{\cal J}_{m \, n \, r \, s} (\Pi) &=~ - \, {N_C}  [ \, 8 {~}_{\dot {~}} 5!
f_{\pi}^5 \, ]^{-1} f_{a \, b}{}^k \, f_{c \, d}{}^l {\rm {Tr}} \Big[ \,
\l_k \l_l \l_h \, \Big] \, \int_0^1 d y ~ y^4 \, \Pi^e {\Hat L}_e {}^h
{\Hat L}_m {}^a {\Hat L}_n {}^b {\Hat L}_r {}^c {\Hat L}_s {}^d  ~~~~. }
\eqno(A.6) $$
where ${\Hat L}_m {}^i \equiv L_m {}^i (y \Pi)$.  Also $f_{a \, b}{}^k$ denotes
the structure constants of the group defined by $ [\l_a , \l_b ] = i f_{a \, b}
{}^k \l_k $.  The effective QCD Lagrangian is simply given
$$ {\cal S}_{eff} ~=~ {\cal S}_{\s} ~+~ {\cal S}_{WZNW}
\eqno(A.7) $$
with $ {\cal S}_{\s}$ defined in (A.3) and $ {\cal S}_{WZNW}$ defined in (A.6).

\noindent
%%%%%%%%%%%%%%%%%%%%%%%%%%%%%%%%%%%%%%%%%%%%%%%%%%%%%%%%%%%%%
{\bf {Appendix B:  Manifest Supersymmetric Formulation of
Kazama-Suzuki \newline ${~~~~~~~~~~~~~~~~~}$ Models and New
(2,2) Superstrings}}
%%%%%%%%%%%%%%%%%%%%%%%%%%%%%%%%%%%%%%%%%%%%%%%%%%%%%%%%%%%%%

In heterotic string theory, one of the well known N = 2 compactification
techniques is give by Kazama-Suzuki models \cite{KS}. An erstwhile
mystery has been, ``How does one find a superfield formulation of
Kazama-Suzuki models?'' Up until now no one has been able to provide
an answer. We now wish to suggest that the missing ingredient seems
to have been the use of the non-minimal scalar multiplet reduced from
4D, N = 1 superspace down to 2D, N = 2 superspace. The reduction itself
is trivial if we introduce the 2D, N = 2 supercovariant derivatives
$D_{\a}$ and their conjugates ${\Bar D}_{\a}$ which satisfy
$$
\Big[ \, D_{\a} , ~ D_{\b} \, \Big\} ~=~ 0 ~~~,~~~
\Big[ \, {\Bar D}_{\a} , ~ {\Bar D}_{\b} \, \Big\} ~=~ 0 ~~~,~~~
\Big[ \, {D}_{\a} , ~ {\Bar D}_{\b} \, \Big\} ~=~ i (\g^c)_{\a \b}
\pa_c ~~~~.
\eqno(B.1) $$
The 2D, N = 2 non-minimal multiplet is now defined by ${\Bar D}^{\a}
{\Bar D}_{\a} \S = 0$. The component fields are defined with a few very
slight modifications (below we use the chiral components)
$$
\eqalign{ {~~~~}
B &\equiv~  \S \, | ~~~,~~~ {\Bar \zeta}_{ \pm} ~\equiv~ {\Bar D}_{\pm} \S
\, | ~~~, ~~~
{\r}_{\pm} ~\equiv~ {D}_{\pm}  \S \, | ~~~,~~~ H ~\equiv~ - i
\, {D}_+ {D}_- \S \, | ~~~,\cr
u &\equiv~ - i \, {\Bar D}_+ D_- \S \, | ~~~,~~~
v ~\equiv~  - i \,{\Bar D}_- D_+ \S \, | ~~~, ~~~
p_{\dpx} ~\equiv~  - i \, {\Bar D}_+ D_+ \S \, | ~~~,~~~\cr
p_{\DM} &\equiv~  - i \, {\Bar D}_- D_- \S \, | ~~~, ~~~
{\Bar \b}_{\pm} ~\equiv~ - i \,  D_+ {\Bar D}_{\pm} D_- \S
\, | ~~~~. }
\eqno(B.2) $$
The complex quantities $u$ and $v$ are the extra components arising
from the dimensional reduction of $p_{\underline a}$ from 4D.
The 2D supersymmetry variations take the forms

$$
\eqalign{ {~~~~~~~~~~~~~~~~}
\d_Q \, B &=~ {\Bar \e}^+ {\Bar \zeta}_+  ~+~ {\Bar \e}^- {\Bar \zeta}_-
{}~+~ \e^+ \r_+  ~+~ \e^- \r_- ~~~~, \cr
\d_Q \, {\Bar \zeta}_+ &=~  i\, \e^+ (\, \pa_{\dpx}  B ~-~ p_{\dpx}
\,)  ~-~ i \ \e^- u   ~~~~,    \cr
\d_Q \, {\Bar \zeta}_- &=~  ~-~ i \, \e^+ v ~+~ i\, \e^- (\, \pa_{\DM}
B ~-~ p_{\DM}  \,)   ~~~~,    \cr
\d_Q \, \r_+  &=~ - i \, \e^- H  ~+~ i\, {\Bar \e}^+  p_{\dpx}
{}~+~ i\, {\Bar \e}^- v  ~~~~,     \cr
\d_Q \, \r_-  &=~  i \, \e^+ H  ~+~ i\, {\Bar \e}^+ u
{}~+~ i\, {\Bar \e}^-  p_{\DM}   ~~~~,     \cr
\d_Q \, u  &=~ \e^+ \b_+ ~-~ {\Bar \e}^- \pa_{\DM} {\Bar \zeta}_+  ~~~~, \cr
\d_Q \, v  &=~ \e^- (\, \pa_{\DM} \r_+ ~-~ \b_- \,)  ~-~ {\Bar \e}^+
\pa_{\dpx} {\Bar \zeta}_- ~~~~,     \cr
\d_Q \, H  &=~  -\, i \, {\Bar \e}^+  (\,  \pa_{\dpx} \r_- ~-~ \b_+
\,)  ~-~ {\Bar \e}^-  \b_-    ~~~~,    \cr
\d_Q \, p_{\dpx}  &=~  \e^+ \pa_{\dpx} \r_+ ~+~ \e^- (\, \pa_{\dpx}
\r_- ~-~ \b_+ \,) ~+~ {\Bar \e}^- \pa_{\dpx} {\Bar \zeta}_- ~~~~,    \cr
\d_Q \, p_{\DM}  &=~ \e^+ \b_- ~+~ \e^- \pa_{\DM} \r_-  ~+~ {\Bar \e}^+
\pa_{\DM} {\Bar \zeta}_+ ~~~~,    \cr
\d_Q \, \b_+  &=~  i\, {\Bar \e}^+ \pa_{\dpx} u ~+~ i {\Bar \e}^- \pa_{\DM}
(\, \pa_{\dpx} B ~-~ p_{\dpx}  \,)   ~~~~,    \cr
\d_Q \, \b_-  &=~ - i\, {\e}^- \pa_{\DM} H ~-~ i {\Bar \e}^+ (\, \pa_{\DM}
\pa_{\dpx} B ~-~ \pa_{\DM} p_{\dpx} ~-~ \pa_{\dpx} p_{\DM} \,)
{}~~~~. }
\eqno(B.3) $$
Finally for the superfield action that should act as the starting point
for the 2D (2,2) Kazama-Suzuki models we propose
$$\eqalign{ {~~~~}
{\cal S}_{KS} ~=~  &\int d^2 \s \,  d^2 \z \, d^2 {\Bar \z} ~ {\Hat \O} (
{\Phi}, \, {\Bar \Phi}; {\S} , \, {\Bar \S} \,)   \cr
& ~+~ \Big[ \int d^2 \s \, d^2 \z ~  {\cal J}_{\rm {I \, J}}
(\Phi) \, ({\Bar D}_+ \S^{\rm I} \,) \, ({\Bar D}_- \S^{\rm J} \,) ~+~
{\rm {h. \, c.}} ~+~ ... ~ \Big]
{}~~~~. }
\eqno(B.4) $$
The terms in the ellipsis represent the introduction of world sheet
2D, N = 1 gauge superfields for the $H$ sub-group in the K-S constructions.
In (B.4) the potential $ {\Hat \O} ( {\Phi}, \, {\Bar \Phi}; {\S} , \,
{\Bar \S} \,)$ is most likely given by (2.7) with $g_{\rm { I \, J}}$
constructed from the Maurer-Cartan forms as in (A.4) and ${\cal J}_{
\rm {I \, J}}(\Phi)$ is given by
$$
{\cal J}_{\rm { I \, J}}(\Phi) ~=~ - \, c_0   f_{\rm { K \, L \, M}}
 \, \int_0^1 d y ~ y^2 \, \Phi^{\rm N} {\Hat L}_{\rm N} {}^{\rm K}
{\Hat L}_{\rm I} {}^{\rm L } {\Hat L}_{\rm J} {}^{\rm  M}  ~~~~.
\eqno(B.5) $$
Here the Maurer-Cartan forms are defined in terms of the chiral superfields
and the group is arbitrary.  However, the final arbiter that determines
these functions is 2D, N = 2 superconformal invariance. This is a topic
to be studied in the future.   Thus, we see for every compact group,
there exist a way to construct a 2D, N = 2 action that possesses manifest
supersymmetry.  Let us emphasize that (2.7) is an explicit construction
that associates with every group manifold with metric $g_{\rm {I \, J}}$
(constructed from the group Maurer-Cartan forms) a special K\" ahler geometry
with a metric whose potential is given by (2.7).  To our knowledge this is
the first observation relating group manifolds to special Kahler geometry
in this manner.  It will be of interest to see if the condition of quantum
superconformal invariance acts as a restriction to the choices considered by
Kazama and Suzuki. Finally, we note that there must exists twisted versions
of the action of (B.5). That is the chiral superfields in (B.5) can be
replaced by twisted chiral superfields, if simultaneously we replace the
complex linear superfields by twisted complex linear superfields, $\Xi$,
(i.e. $\S \to \Xi$ where $\Xi$ satisfies $ {\Bar D}_+ D_- \Xi = 0$).

Let us be explicit, we expect a subclass of the actions of (B.4) to
describe a fundamentally {\underline {new}} class of 2D, N = 2
superstrings. As long ago as 1989, we reported that at the level
of superfields\footnote{See S.J.Gates, Jr., R. Oerter and L. Lu,
Phys. Lett. 218B (1989) 33.} there were at least three different
N = 2 superstring actions. One of these, which actually has an N = 4
rigid supersymmetry (one chiral plus one twisted chiral
multiplet) is known to permit a non-trivial axion background
unlike the other two version. However, the axion occurs as
the second derivative of a potential. Our new theories are not
subject to this constraint. So we believe with (B.4) we have yet
again increased the number of known 2D, N = 2 superstrings. These
new N = 2 superstrings are associated with different choices of
auxiliary fields.  So even for string theory we have evidence that
auxiliary fields matter...a point totally absent in superconformal
field theory.

\newpage
%%%%%%%%%%%%%%%%%%%%%%%%%%%%%%%%%%%%%%%%%%%%%%%%%%%%%%%%%%%%%%%%%%%%
%\sect{References}
%%%%%%%%%%%%%%%%%%%%%%%%%%%%%%%%%%%%%%%%%%%%%%%%%%%%%%%%%%%%%%%%%%%%

\end{document}